\documentclass[preprint,prd,nofootinbib,tightenlines,amsmath]{revtex4}
\usepackage{enumerate}
\usepackage{multirow}

\usepackage{epsfig}
\usepackage{graphics,color}      
\usepackage{verbatim}
\usepackage{amsmath}    
\catcode`\@=11
\def\sla@#1#2#3#4#5{{%
 \setbox\z@\hbox{$\m@th#4#5$}%
 \setbox\tw@\hbox{$\m@th#4#1$}%
 \dimen4\wd\ifdim\wd\z@<\wd\tw@\tw@\else\z@\fi
 \dimen@\ht\tw@
 \advance\dimen@-\dp\tw@ \advance\dimen@-\ht\z@
 \advance\dimen@\dp\z@
 \divide\dimen@\tw@ \advance\dimen@-#3\ht\tw@
 \advance\dimen@-#3\dp\tw@ \dimen@ii#2\wd\z@
 \raise-\dimen@\hbox to\dimen4{%
 \hss\kern\dimen@ii\box\tw@\kern-\dimen@ii\hss}%
 \llap{\hbox to\dimen4{\hss\box\z@\hss}}}}

\def\slashed#1{%
 \expandafter\ifx\csname sla@\string#1\endcsname\relax
{\mathpalette{\sla@/00}{#1}}
\fi}
\def\declareslashed#1#2#3#4#5{%
 \expandafter\def\csname sla@\string#5\endcsname{%
#1{\mathpalette{\sla@{#2}{#3}{#4}}{#5}}}}
 \catcode`\@=12
\declareslashed{}{/}{.08}{0}{D}
 \declareslashed{}{/}{.1}{0}{A}
 \declareslashed{}{/}{0}{-.05}{k}
 \declareslashed{}{/}{.1}{0}{\partial}
 \declareslashed{}{\not}{-.6}{0}{f}

\def\lsim{\mathrel {\vcenter {\baselineskip 0pt \kern 0pt
    \hbox{$<$} \kern 0pt \hbox{$\sim$} }}}
\def\gsim{\mathrel {\vcenter {\baselineskip 0pt \kern 0pt
    \hbox{$>$} \kern 0pt \hbox{$\sim$} }}}

\newcommand{\bea}{\begin{eqnarray}}
\newcommand{\eea}{\end{eqnarray}}

\begin{document}

\baselineskip=15pt
\preprint{}

\title{CP violating anomalous couplings in $W\gamma$ and $Z\gamma$ production at the LHC}

\author{Sally Dawson$^1$, Sudhir Kumar Gupta$^2$ and German Valencia$^3$}

\email{dawson@bnl.gov,sudhir.gupta@monash.edu,valencia@iastate.edu}

\affiliation{$^1$ Department of Physics, Brookhaven National Laboratory, Upton, NY 11973, USA\\
$^2$ ARC Centre of Excellence for Particle Physics at the Terascale, School of Physics, Monash University, Melbourne, Victoria 3800 Australia\\
$^3$ Department of Physics, Iowa State University, Ames, IA 50011, USA.}

\date{\today}

\vskip 1cm
\begin{abstract}

The ATLAS and CMS collaborations have recently published new limits on CP conserving anomalous couplings from the $W\gamma$ and $Z\gamma$ production processes. We study the corresponding limits that can be placed on the CP violating anomalous couplings $\tilde\kappa_\gamma$ and $h^1_{ \gamma,Z}$ at the LHC. We find that the process $pp\rightarrow W\gamma$ at 14 TeV can place the $95\%$ CL limit  $\mid {\tilde \kappa}_\gamma\mid\lsim 0.05$ with 10 fb$^{-1}$ which is comparable to the existing LHC bound on the CP conserving anomalous couplings $\kappa_\gamma$. Similarly, the process $pp\rightarrow Z\gamma$ can place the $95\%$ CL limits 
$\mid h^1_{ \gamma}\mid\lsim 20$ and $\mid h^1_{ Z}\mid\lsim 40$ respectively.  None of these limits is derived from a truly CP-odd observable so that it is not possible to separate the effects of the CP violating anomalous couplings from the rest.
\end{abstract}

\pacs{PACS numbers: }

\maketitle

\section{Introduction}

The Large Hadron Collider (LHC) has already delivered an integrated 
luminosity of more than 5~fb$^{-1}$ per experiment during its 7 TeV run, 
and another 20~fb$^{-1}$ at 8 TeV, and is well on its way to testing the 
Standard Model (SM) in detail. One of these tests involves a measurement 
of the triple gauge boson couplings $WW\gamma$ and $ZZ\gamma$ through 
their contribution to $W\gamma$ and $Z\gamma$ production. Early reports 
of these measurements include one from ATLAS with 4.6 fb$^{-1}$ 
\cite{ATLAS:2012mec} as well as older ATLAS \cite{Aad:2012mr, Aad:2011tc} and CMS \cite{Martelli:2012np, Chatrchyan:2011rr} results with 2010 data with an integrated luminosity of 1.02 fb$^{-1}$.

These measurements are used to constrain physics beyond the SM by parametrizing the $WW\gamma$ and $ZZ\gamma$ vertices with anomalous
 couplings \cite{Hagiwara:1986vm,Gounaris:2000tb} and placing limits on the values of these couplings. Although the anomalous coupling parametrization is completely general, it is a common practice to consider only the subset of couplings that conserve charge conjugation and parity (and thus CP). In this paper we revisit the study of the subset of these couplings that violate CP along the lines of our previous analysis for the Tevatron \cite{Dawson:1996ge}.

The main motivation for this study is that our understanding of CP violation remains incomplete and constraining the CP violating anomalous couplings from the experimental perspective it is a simple extension of existing studies.

From measurements of the cross-section alone it is not possible to distinguish between the effects of the CP conserving or CP violating anomalous couplings,
although we will show that the $Z\gamma$ cross section is quite sensitive to CP violating couplings. 
In order to separate the two it is necessary to study a CP violating observable. 
To this end we consider naive-T odd triple product correlations \cite{tprods} as we did in Ref. \cite{Dawson:1996ge}. 
We find that these correlations are sensitive  to the presence of CP violating
interactions in $W\gamma$ production, but do not uniquely require  CP violation.
The sensitivity to CP violating anomalous couplings in $W\gamma$ production is discussed in Section \ref{wg}. 
In the $Z\gamma$ process, the asymmetries we construct do single out CP violation,  but the sensitivity at the LHC
 is low.

\section{$pp \to W^\pm\gamma \to \ell^\pm \nu \gamma$}
\label{wg}

We begin by examining new physics effects that can affect the $WW\gamma$ vertex, and potentially be observed in 
$pp \to W^\pm\gamma \to \ell^\pm \nu \gamma$.
The anomalous couplings that may affect this process can be found in Ref.~\cite{Hagiwara:1986vm},
\begin{eqnarray}
\frac{{\cal L}_{WWV}}{g_{WWV}} &=& i
g^V_1(W^{\dagger}_{\mu\nu}W^{\mu}V^{\nu} -
W^{\dagger}_{\mu}V_{\nu}W^{\mu\nu}) \nonumber \\
&& + i \kappa_V W^{\dagger}_{\mu}W_{\nu}V^{\mu\nu}
+\frac{i\lambda_V}{m^2_W}W^{\dagger}_{\lambda\mu}W^{\mu}_{\nu}V^{\nu\lambda}
 -g^V_4 W^{\dagger}_{\mu}W_{\nu}(\partial^{\mu}V^{\nu}+\partial^{\nu}V^{\mu})
\nonumber \\
&&+g^V_5 \epsilon^{\mu\nu\rho\sigma}(W^{\dagger}_{\mu}
(\overrightarrow{\partial_{\rho}}-\overleftarrow{\partial_{\rho}})W_{\nu})V_{\sigma}
+i \tilde{\kappa}_V W^{\dagger}_{\mu}W_{\nu}\tilde{V}^{\mu\nu}
+\frac{i\tilde{\lambda}_V}{m^2_W}W^{\dagger}_{\lambda\mu}W^{\mu}_{\nu}\tilde{V}^{\nu\lambda},
\label{hagiwwvdef}
\end{eqnarray}
where $V^\mu$ stands for either the photon or $Z$ field, $W^\mu$ is the $W^-$ field, 
$W_{\mu\nu}=\partial_\mu W_\nu-\partial_\nu W_\mu$, $V_{\mu\nu}=\partial_\mu V_\nu-\partial_\nu V_\mu$, 
$\tilde{V}_{\mu\nu}=\frac{1}{2}\epsilon_{\mu\nu\rho\sigma}V^{\rho\sigma}$.  
The  gauge couplings are $g_{WW\gamma}=-e$ and
$g_{WWZ}=-e\cot\theta_W$.
In the SM, $g^V_1=\kappa_V=1$ and all the others are zero.  The couplings 
in Eq. \ref{hagiwwvdef} that violate CP are ${\tilde \kappa}_V$ and ${\tilde \lambda}_V$.

The CP conserving couplings present in Eq.~\ref{hagiwwvdef} that  conserve 
both C and P and affect $W\gamma$ production have been constrained  by ATLAS 
and CMS. The most recent result \cite{ATLAS:2012mec} gives an allowed 95\% 
CL interval (without assuming an arbitrary form factor or any relation between the couplings) of

\begin{eqnarray}
-0.135 \leq & \kappa_\gamma -1 & \leq 0.190 \nonumber \\
-0.152 \leq & \lambda_\gamma & \leq 0.146 \nonumber \\
-0.078 \leq & \kappa_Z -1 & \leq 0.092 \nonumber \\
-0.074 \leq & \lambda_Z & \leq 0.073 \nonumber \\
-0.373 \leq & g_1^Z-1 & \leq 0.562\, ,
\label{atlaslimits}
\end{eqnarray}
which are comparable to older Tevatron results \cite{tevlimitW} and a bit worse than the LEP limits \cite{Alcaraz:2006mx}. 

The P and CP violating couplings in Eq.~\ref{hagiwwvdef} that enter into the $W\gamma$ production process are $\tilde\kappa_{\gamma}$ and 
$\tilde\lambda_{\gamma}$~\cite{Hagiwara:1986vm, Baur:1989gk}. We limit our study to the former as the latter originates in a higher dimension operator \cite{Appelquist:1994qz}. The corresponding couplings $\tilde\kappa_{Z}$ and $\tilde\lambda_{Z}$ have been recently studied for the  LHC. Ref.~\cite{Han:2009ra} finds that the LHC can reach a $5\sigma$ sensitivity of 0.1 for both of them with 100~fb$^{-1}$ by measuring the observable
\begin{equation}
sgn((p_{\ell^+}-p_{\ell^-}).\hat{z})\sin^{-1}((p_{\ell^+}\times p_{\ell^-}).\hat{z})
\label{hancpodd}
\end{equation}
in the $W^+W^-$ production process. This process has the advantage of permitting the construction of a genuinely CP-odd observable such as Eq.~\ref{hancpodd}.
The coupling $\tilde{\lambda}_Z$ has also been studied in this context in Ref.~\cite{Kumar:2008ng} in $WZ$ production by measuring the observable
\begin{equation}
\Delta = \int d \sigma sgn(k_Z^z)sgn(p_\ell \times k_Z)^z\, .
\end{equation}
They find a significantly improved sensitivity of $|\tilde{\lambda}_Z|\sim 0.002$ at the $5\sigma$ level with 100~fb$^{-1}$. 

In Ref.~\cite{Dawson:1996ge} we gave an analytic result for the CP-violating interference term (linear in $\tilde\kappa_\gamma$) in the parton level process $q \bar{q}\to \ell^\pm\nu \gamma$. That result, (Eq.~3 of Ref.~\cite{Dawson:1996ge}), suggests that
 the T-odd observables that are better suited for constraining  $\tilde\kappa_\gamma$ are,
\begin{eqnarray}
{\cal O}_W &=&  \left( {\vec p}_\gamma \times {\vec p}_{beam} \right) .  {\vec p}_\ell \nonumber \\
{\cal O}_\gamma &=&  {\vec p}_\gamma . {\vec p}_{beam} \  {\cal O}_W \nonumber \\
{\cal O}_\ell &=&  {\vec p}_\ell. {\vec p}_{beam}\  {\cal O}_W\, .
\label{tripprods}
\end{eqnarray}
In Ref.~\cite{Dawson:1996ge} we used ${\cal O}_W$ to find that the Tevatron could place the constraint $|\tilde\kappa_\gamma| \lsim 0.1$ at 95\% CL with an integrated luminosity of 10~fb$^{-1}$. This observable does not work at the LHC, however,
 where the identical protons of the initial $pp$ state require a symmetrization of the beam direction. Instead, we must use one of the other two operators which are quadratic in the beam direction. We can use each  of the operators in Eq.~\ref{tripprods} to construct integrated asymmetries defined by
\begin{equation}
{\cal A}_{W,\ell,\gamma} \equiv \left(\sigma({\cal O}_{W,\ell,\gamma} >0)-\sigma({\cal O}_{W,\ell,\gamma} <0)\right)\, .
\label{astriprods}
\end{equation}
The numerical results reported in the Tables in the Appendix use the notation, for each operator, 
$\sigma^+\equiv\sigma({\cal O}_{W,l,\gamma}>0), \quad {\hbox{etc.}}$

A further complication with respect to the Tevatron study of Ref.~\cite{Dawson:1996ge} 
 is that at the LHC the $W^+\gamma$ and $W^-\gamma$ production processes are not CP conjugates of each other. For this reason the T-odd observables 
 ${\cal O}_{\gamma}$  and ${\cal O}_{\ell}$ from the $W^\pm \gamma$ processes cannot be combined to form CP-odd observables. The net result is that a non-zero result cannot be interpreted as conclusive evidence for CP violation. To obtain conclusive evidence for CP violation at the LHC, one would have to study the $W^+W^-$ production process, which is self conjugate. This would not necessarily single out a particular anomalous coupling, but would provide a process in which the CP properties of the final state can be used to construct truly CP-odd observables \cite{tprods,Han:2009ra}.

We begin by implementing the anomalous couplings $\kappa_\gamma$ and 
${\tilde \kappa}_\gamma$ in  {\tt MadGraph 5} \cite{madgraph} using {\tt FeynRules} \cite{Christensen:2008py}. To test our implementation we first repeat the Tevatron analysis of Ref.~\cite{Dawson:1996ge} for the process $p{\bar p}\to W^\pm \gamma \to l^\pm \nu_l \gamma$ at $\sqrt{S}= 2$ TeV for the couplings ${\tilde\kappa}_\gamma$ and ${\kappa}_\gamma$ respectively. We use the parton distribution functions {\tt CTEQ6L1} 
evaluated at a scale $\mu_R = \mu_F = \sqrt{M^2_W + 
p^2_{T_\gamma}}$, and impose the cuts: $p_{T_l} > 20$ GeV, $p_{T_\gamma} > 10$ GeV, $\left|\eta_l\right| < 3.0$, 
$\left|\eta_\gamma\right| < 2.4$, $\Delta R_{l\gamma} > 0.7$, $\slashed p_T > 20$ GeV, and $M_T(l \gamma, \slashed p_T) > 90$ GeV. We proceed to generate $10^{7}$ events to measure ${\cal A}_W$ 
and present raw numbers in Table~\ref{tab:ktila_tev_ow} in the Appendix.

\begin{figure}[htb]  
\centerline{\includegraphics[angle=0, width=1.2\textwidth]{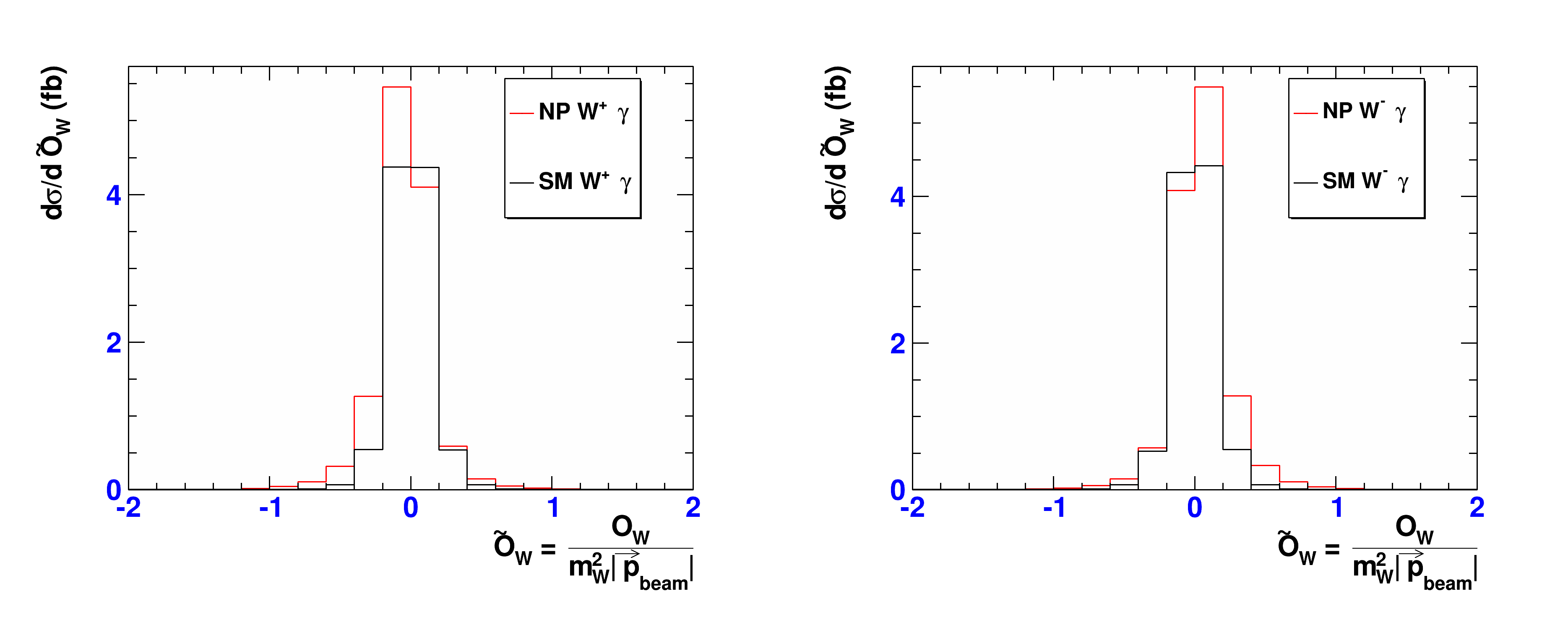}}
\vspace{-0.6cm}
\caption{\sf\small ${\cal O}_W$ distributions at the Tevatron for the SM (Black) and NP (Red) with $\tilde\kappa_\gamma = 0.3$. Cuts are same as in Table~\ref{tab:ktila_tev_ow}.
Left plot: $p {\bar p}\to W^+ \gamma \to l^+ \nu_l \gamma$, Right plot: $p {\bar p} \to W^- \gamma \to l^- \nu_l \gamma$.}
\label{fig:otev}
\end{figure}

Our results for the cross-sections and asymmetry corresponding to ${\cal O}_W$ for $\sqrt{S}=$2~TeV at the Tevatron are,
\begin{eqnarray}
\sigma^+_{SM} &=& \sigma^-_{SM} \ =\ 200~{\rm fb} \nonumber \\
{\cal A}_W^+ &=& - {\cal A}_W^- = -158 ~\tilde{\kappa}_\gamma ~{\rm fb},
\end{eqnarray}
in rough agreement with Ref.~\cite{Dawson:1996ge}. The difference, of about 25\%,  we attribute mostly to the different parton distribution functions (a now obsolete set of PDF's was used in 1996). In addition, the older results were obtained with a significantly smaller number of events. The different sign in the asymmetry 
relative to Ref.~\cite{Dawson:1996ge} is due to a sign difference in the definition of the triple product. Our results also confirm that the CP conserving coupling $\kappa_\gamma$ does not induce the T-odd observable without the inclusion of absorptive phases.  The distributions of ${\cal O}_W$ for $W^\pm\gamma$ at the Tevatron
are shown in Fig. \ref{fig:otev} for the rather large value ${\tilde \kappa}_\gamma = 0.3$, which we use to make the asymmetry clearly visible. 

We proceed to study the case of the LHC  at $\sqrt{S}=$14 TeV with {\tt MadGraph 5} using {\tt CTEQ6L1} pdf's evaluated at a scale $\mu_R = \mu_F = \sqrt{M^2_W + 
p^2_{T_\gamma}}$ and imposing the set of cuts:  $p_{T_l} > 20$ GeV, $p_{T_\gamma} > 20$ GeV, $\left|\eta_l\right| < 2.5$, 
$\left|\eta_\gamma\right| < 2.5$, $\Delta R_{l\gamma} > 0.7$, $\slashed p_T > 20$ GeV, and $M_T(l \gamma, \slashed p_T) > 90$~GeV. 
\footnote{
We define 
$
M_T(l\gamma,\slashed p_T)
\equiv  \biggl(
\sqrt{
m_{\l \gamma}^2+
\mid
{\vec{p}_{T_l}}
+{\vec {p}_{T_\gamma}}
\mid^2}
+p_{T_{\nu}}
\biggr)^2
-\mid 
{\vec {p}_{T_l}}
+{\vec {p}_{T_\gamma}}
+{\vec {p}_{T_\nu}}
\mid^2$. The numerical results are quite sensitive to this choice.
}
For comparison purposes we keep both $\kappa_\gamma$ and $\tilde\kappa_\gamma$ and construct asymmetries based on the three T-odd correlations of Eq.~\ref{astriprods}. 
Our results after generating $10^6$ events are shown in Table~\ref{tab:ktila_lhc_ow} of the Appendix.

The results confirm that there is no asymmetry linear in the beam momentum, as required by the symmetry of the initial $pp$ state, thus ${\cal A}_W=0$ in all cases. The two asymmetries quadratic in the beam momentum, ${\cal A}_{\gamma}$ and  ${\cal A}_{\ell}$, 
are very similar numerically (although they have opposite signs) and they vanish for the case of CP conserving physics without absorptive
 phases (SM and NP2 cases in Table~\ref{tab:ktila_lhc_ow}).\footnote{
 We work at tree level in the SM.} We can use our simulation to write approximate expressions for $pp\rightarrow W^\pm \gamma \rightarrow l^\pm \nu_l \gamma$ cross-sections and asymmetries,
\begin{eqnarray}
\sigma(W^+\gamma) &\approx & \left( 659.6 + 120.2 ~\kappa_\gamma + 4558.3~ \kappa^2_\gamma -  1.3 ~{\tilde\kappa}_\gamma + 4573.3 ~ {\tilde \kappa}^2_\gamma\right) {\rm ~fb}\nonumber \\
\sigma(W^-\gamma) &\approx & \left(549.6 + ~66.3 ~\kappa_\gamma + 2986.7~ \kappa^2_\gamma -  5.5 ~{\tilde\kappa}_\gamma + 3065.0 ~ {\tilde \kappa}^2_\gamma\right) {\rm ~fb}\nonumber \\
{\cal A}_\gamma (W^+\gamma) &=& -{\cal A}_\ell(W^+\gamma)  \ \approx \ 462~ \tilde\kappa_\gamma
\nonumber \\
{\cal A}_\gamma (W^-\gamma) &=& -{\cal A}_\ell(W^-\gamma)  \ \approx \ 452~ \tilde\kappa_\gamma \, .
\label{wgres}
\end{eqnarray}

\begin{figure}[htb]
\centerline{\includegraphics[angle=0, width=1.2\textwidth]{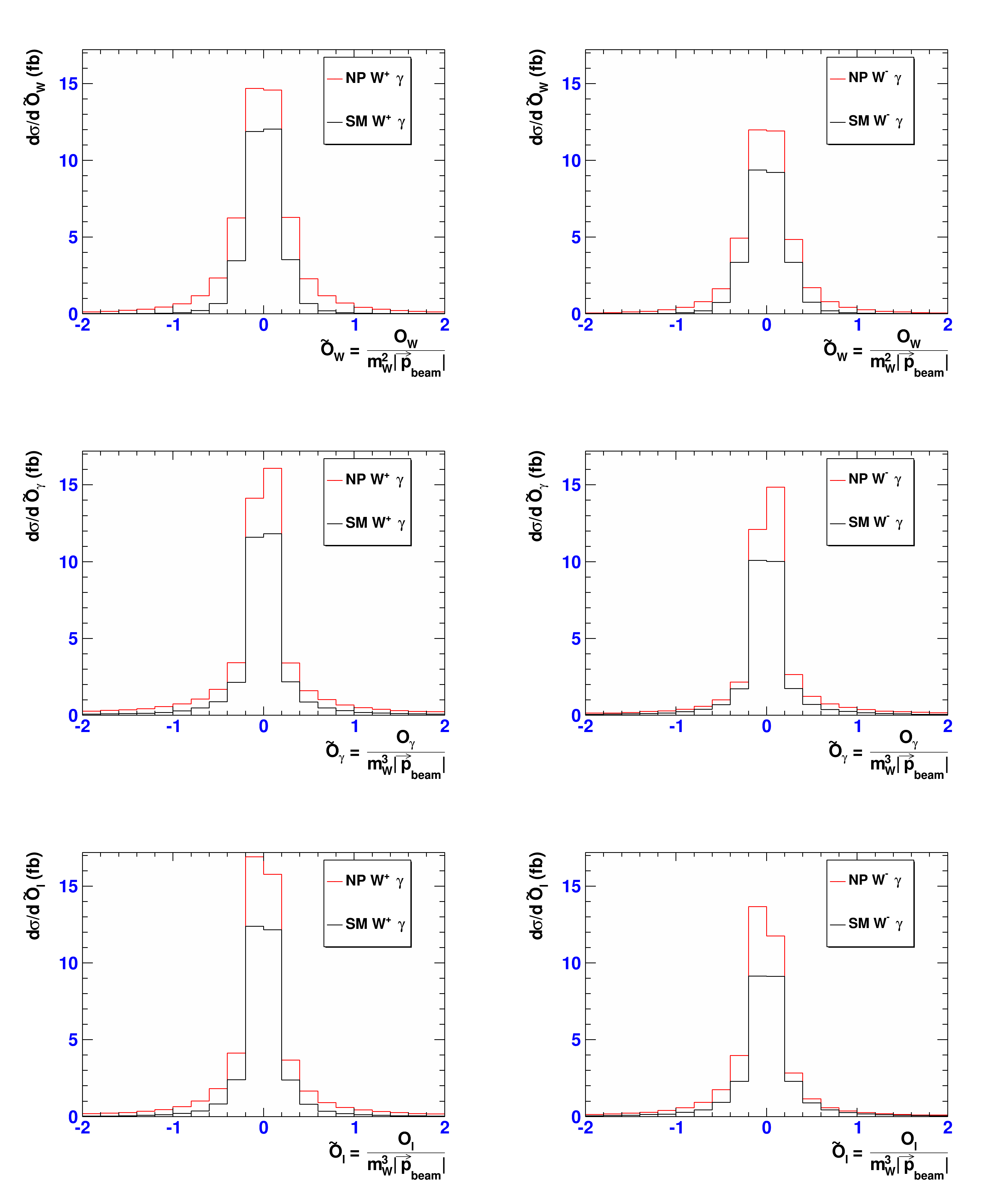}}
\vspace{-0.6cm}
\caption{\sf\small ${\cal O}_W$ (top), ${\cal O}_\gamma$ (middle) and ${\cal O}_\ell$ (bottom) distributions at the LHC 
with $\sqrt{S}=14$ TeV for the SM (Black) and NP (Red) with $\tilde\kappa_\gamma = 0.3$. Cuts are same 
 as in Table~\ref{tab:ktila_lhc_ow}. Left plot: $p p\to W^+ \gamma \to l^+ \nu_l \gamma$, Right plot: $p p\to W^- \gamma \to l^- \nu_l \gamma$.}
\label{fig:olhc}
\end{figure}

These results show a small interference between the new physics and the SM for the case of $\kappa_\gamma$, which results in a sensitivity to this coupling that arises primarily from its quadratic contributions to the cross-section. Notice that the interference terms between $\tilde\kappa_\gamma$ and the SM are consistent with zero within our statistical error (see Table~\ref{tab:ktila_lhc_ow}). This is of course expected as there can be no interference between the CP violating coupling and the SM amplitude unless one goes beyond tree level and allows absorptive phases to occur. 
The distributions of the operators of Eq. \ref{tripprods} are shown in Fig. \ref{fig:olhc}, from which the small asymmetries
of Eq. \ref{wgres} can be observed.  In particular we see that the distribution for ${\cal O}_W$ is symmetric within  statistical error in agreement with our expectation that this operator cannot produce an asymmetry at LHC. At the same time the asymmetries associated with ${\cal O}_{\gamma,\ell}$ are non-zero and have opposite signs.
 
\begin{figure}[htb]
\centerline{\includegraphics[angle=0, width=1.2\textwidth]{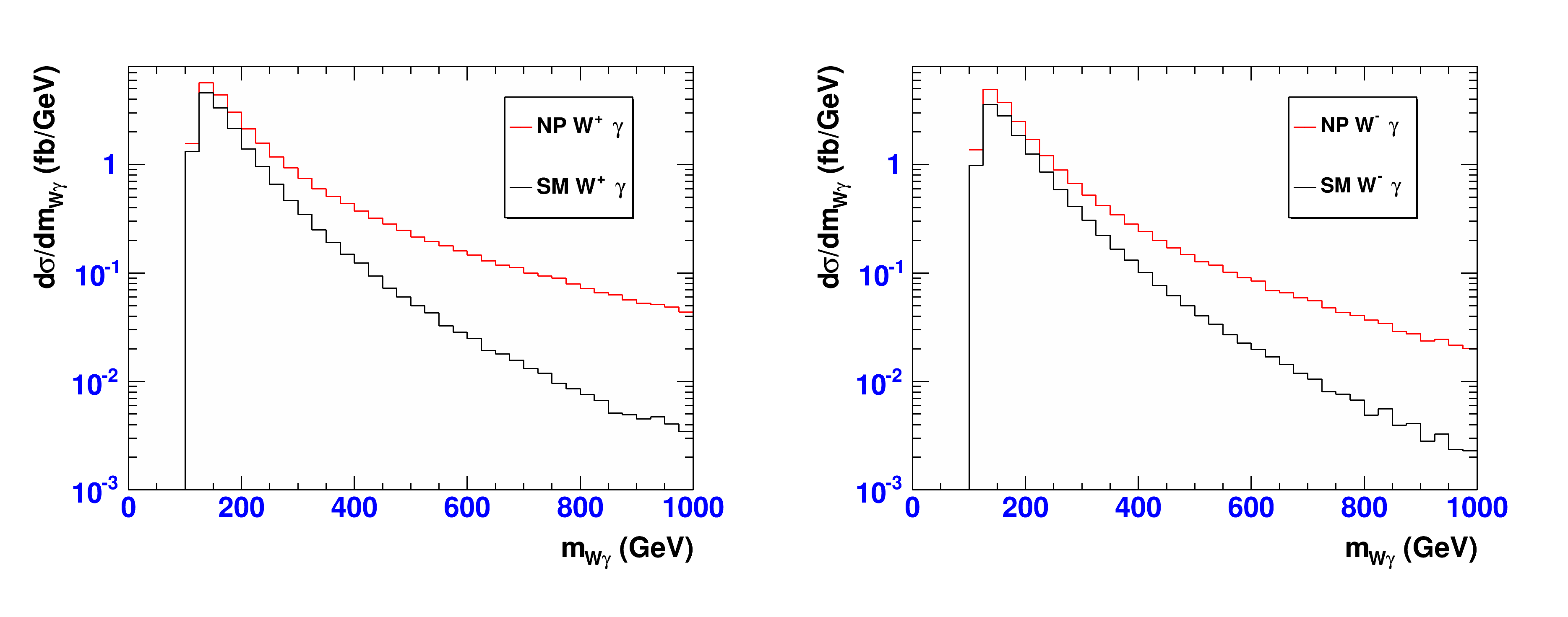}}
\vspace{-0.6cm}
\caption{\sf\small $d\sigma/dm_{W\gamma}$ plots at the LHC
with $\sqrt{S}=14$ TeV
 for the SM (black) and NP (red) with $\tilde\kappa_\gamma = 0.3$. Cuts are same
 as in Table~\ref{tab:ktila_lhc_ow}. Left plots : $p p\to W^+ \gamma \to l^+ \nu_l \gamma$, Right plots: $p p\to W^- \gamma \to l^- \nu_l \gamma$.}
\label{fig:amwalhc}
\end{figure}

\begin{figure}[htb]
\centerline{\includegraphics[angle=0, width=1.2\textwidth]{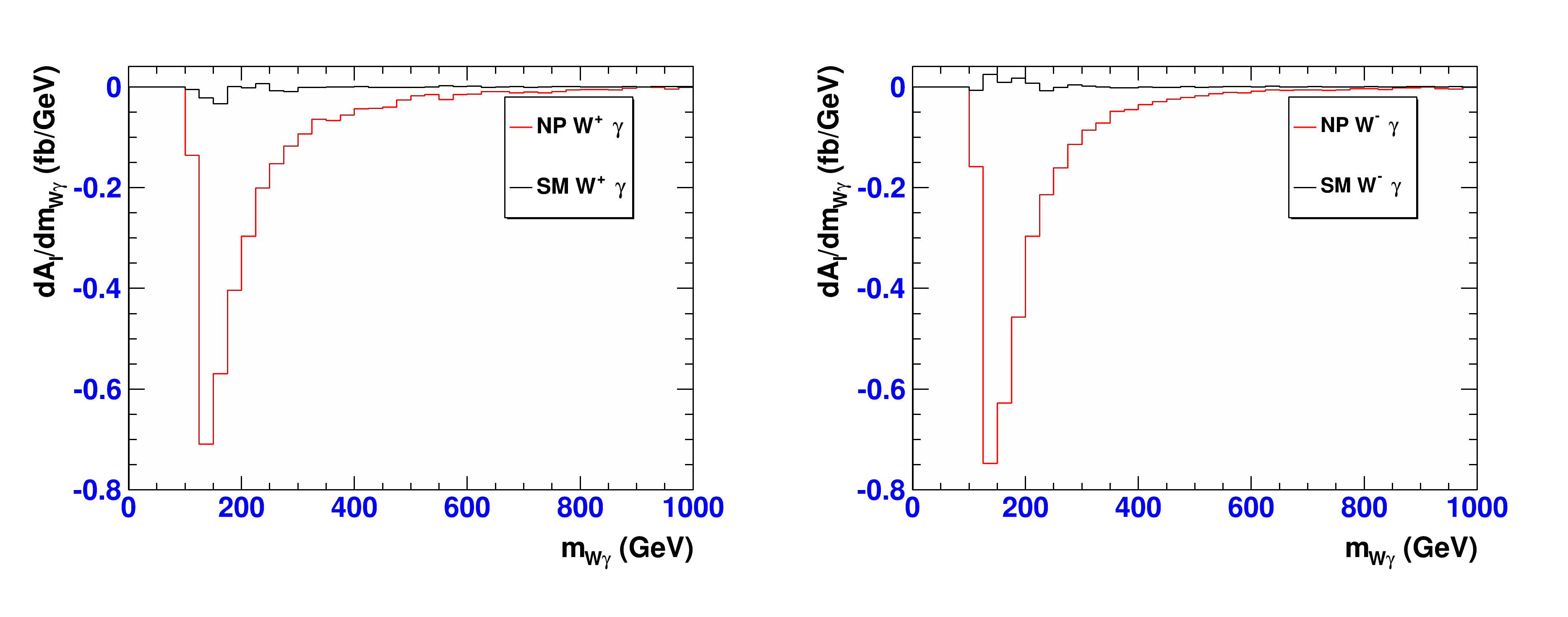}}
\vspace{-0.6cm}
\caption{\sf\small $d{\cal A}/dm_{W\gamma}$ plots for the operator ${\cal O}_\ell$ at the LHC
with $\sqrt{S}=14$ TeV
 for the SM (black) and NP (red) with $\tilde\kappa_\gamma = 0.3$. Cuts are same
 as in Table~\ref{tab:ktila_lhc_ow}. Left plots : $p p\to W^+ \gamma \to l^+ \nu_l \gamma$, Right plots: $p p\to W^- \gamma \to l^- \nu_l \gamma$.}
\label{fig:xmwalhc}
\end{figure}

The invariant mass distributions are shown in Fig. \ref{fig:amwalhc}. We see that the importance of the new physics increases with $m_{W\gamma}$ as expected. The value of $\tilde\kappa_\gamma\sim 0.3$ illustrated in this figure corresponds to a rather low new physics scale around 1~TeV for a dimensionless $\alpha_{13}\sim 10$ \cite{Dawson:1996ge} and therefore the large enhancement in the distribution beyond $M_{W\gamma}\sim 500$~GeV is not realistic. We can nonetheless use these distributions to 
construct the asymmetries corresponding to ${\cal O}_l$, shown in Fig. \ref{fig:xmwalhc}. Interestingly the asymmetries are largest at the low values of invariant mass, $M_{W\gamma}$, where the effective Lagrangian description is more robust.

From Eq. \ref{wgres}  we can estimate the statistical sensitivity, $S$, for a given integrated luminosity, ${\cal L}$, using
\begin{eqnarray}
\frac{(\sigma-\sigma_{SM})}{\sqrt{\sigma_{SM}}}\sqrt{\cal L} = S &{\rm ~or}&
\frac{\cal A}{\sqrt{\sigma_{SM}}}\sqrt{\cal L} = S\, .
\end{eqnarray}
For example, the $95\%$ CL limits for the sensitivity attained from cross-section measurements with 10~fb$^{-1}$ are 
\begin{eqnarray}
-0.07 \lsim \kappa_\gamma \lsim 0.05  && {\rm ~from~}\sigma(W^+\gamma) \nonumber \\
-0.08 \lsim \kappa_\gamma \lsim 0.06  && {\rm ~from~}\sigma(W^-\gamma) \nonumber \\
-0.06 \lsim \tilde\kappa_\gamma \lsim 0.06  && {\rm ~from~}\sigma(W^+\gamma) \nonumber \\
-0.07 \lsim \tilde\kappa_\gamma \lsim 0.07  && {\rm ~from~}\sigma(W^-\gamma).
\label{mad14sensitivity}
\end{eqnarray}
Our estimate for the sensitivity to $\kappa_\gamma$ is in rough agreement with the actual limit obtained by ATLAS with 4.6~fb$^{-1}$ quoted in Eq.~\ref{atlaslimits}. Notice that the results  for $\tilde\kappa_\gamma$ are very similar, and that this can already be inferred from Eq.~\ref{wgres} where the main difference between the two couplings is due to a small interference between the CP conserving $\kappa_\gamma$ and the SM. 

We can also constrain the CP violating coupling $\tilde\kappa_\gamma$ with the T-odd asymmetries. At $95\%$ CL for 10~fb$^{-1}$ we find 
\begin{eqnarray}
\mid \tilde\kappa_\gamma \mid \lsim 0.03,
\end{eqnarray}
with all the asymmetries studied offering roughly the same sensitivity. Unlike the Tevatron, the LHC does not allow us to single out the $WW\gamma$ CP violating anomalous couplings with the study of the T-odd correlations.  CP conserving anomalous couplings, in conjunction with absorptive phases (which occur at one loop and beyond), also generate the T-odd asymmetries. For example, we have confirmed numerically that ${ \kappa}_\gamma$ with an additional absorptive phase  generates  non-zero  values of ${\cal A}_{\gamma}$ and ${\cal A}_{\ell}$. To separate the two cases would require the study of $\bar p\bar p\to W^\mp\gamma$ reactions as well. Nevertheless, the T-odd correlations have a slightly better sensitivity to $\tilde\kappa_\gamma$ than the cross-sections. Finally, it is worth recalling that indirect constraints on CP violating anomalous couplings are typically much better. For example, 
the claimed sensitivity of EDM experiments being $\mid {\tilde \kappa}_\gamma\mid < 5\times 10^{-5}$ \cite{Marciano:1986eh}. As usual, indirect constraints on new physics are complementary to direct high energy constraints and not a substitute for them.

We now turn our attention to the importance of QCD corrections for these processes and their influence on the expected sensitivities. It is known that the lowest order (LO) $W\gamma$ cross-sections we have considered thus far, exhibit a large dependence on the QCD scales $\mu_R$  and 
$\mu_F$\footnote{The choice of PDF set is expected to be much less 
important than the choice of scale, since this is a $q{\overline q}$ 
initiated process, where the PDFs are well determined and the CTEQ, NNPDF,
and MSTW PDFs are in good agreement at $Q^2\sim M_W^2$.}. This effect was
 observed in early studies of anomalous couplings at the Tevatron~\cite{Baur:1993ir}. 
 To quantify the effect of QCD corrections on the sensitivities we have estimated, we implemented  the anomalous coupling  ${\tilde \kappa}_\gamma$
  into {\tt MCFM 6.2} \cite{mcfm} and repeated our calculations for the LHC with $\sqrt{S}= 14$~TeV both at LO and NLO QCD. For the NLO calculations we use the CTEQ6.6M parton  distribution functions. Our results for two different values of $\kappa_\gamma$ and $\tilde\kappa_\gamma$ are displayed in Table~\ref{tab:mcfm_wga_lhc_14}. Comparison of these numbers with those in Table~\ref{tab:ktila_lhc_ow} reveals that
\begin{enumerate}

\item The cross-section results obtained with {\tt MCFM} at LO are within 5\% of the results obtained with {\tt MadGraph}, and these differences are larger at larger $p_T$. Both of these observations are consistent with previous studies \cite{cmswgstudy}.

\item The differences between the two LO simulations depend on the value of the T-odd correlations so that they are magnified in some of the asymmetries, being as large as 20\% for ${\cal A}_\gamma(W^-\gamma)$ but remaining in the 5\% range for ${\cal A}_\gamma(W^+\gamma)$.

\item The NLO cross-sections are significantly larger than the LO cross-sections, again in agreement with previous observations. This K- factor appears to be larger for the SM than for the NP contributions.

\end{enumerate}
  
The corresponding sensitivities estimated from the {\tt MCFM} simulations at the $95\%$ CL  with 10~fb$^{-1}$ are 
\begin{eqnarray}
 {\rm ~from~}\sigma(W^+\gamma)\ \  -0.05 \lsim \kappa_\gamma \lsim 0.08   {\rm ~LO,} && -0.1 \lsim \kappa_\gamma \lsim 0.05 {\rm ~NLO}
\nonumber \\
 {\rm ~from~}\sigma(W^-\gamma)\ \ -0.08 \lsim \kappa_\gamma \lsim 0.07  {\rm ~LO,} && -0.05 \lsim \kappa_\gamma \lsim 0.13  {\rm ~NLO} \nonumber \\
{\rm ~from~}\sigma(W^+\gamma)\ \ -0.06 \lsim \tilde\kappa_\gamma \lsim 0.06  {\rm ~LO,}  && -0.08 \lsim \tilde\kappa_\gamma \lsim 0.06  {\rm ~NLO}  \nonumber \\
{\rm ~from~}\sigma(W^-\gamma)\ \ -0.07 \lsim \tilde\kappa_\gamma \lsim 0.07 {\rm ~LO,} &&  -0.08 \lsim \tilde\kappa_\gamma \lsim 0.09 {\rm ~NLO} .
\label{mcfm14sensitivity}
\end{eqnarray}
From the asymmetries at $95\%$ CL for 10~fb$^{-1}$ we find sensitivities in the range
\begin{eqnarray}
\mid \tilde\kappa_\gamma \mid \lsim 0.03 -0.05,
\end{eqnarray}
depending on whether we use ${\cal A}(W^\pm\gamma)$ at LO or NLO. Comparing these results with Eq.~\ref{mad14sensitivity} we see that although there is some uncertainty from the different MC and from NLO effects, the constraints that can be placed on the CP violating coupling $\tilde\kappa_\gamma$ are roughly equal to those that can be placed on its CP conserving counterpart.
  
We end this section presenting results for $\sqrt{S}=7$~TeV, imposing in this case the appropriate set of cuts given by 
\begin{eqnarray}
p_{T\ell} > 20~{\rm GeV}, && p_{T\gamma} > 10~{\rm GeV} \nonumber \\
|\eta_\ell| < 2.5, && |\eta_\gamma| < 2.4 \nonumber \\
\Delta R_{\ell \gamma} > 0.7 && |\slashed{p}_T| > 25~{\rm GeV} \nonumber \\
M_T(\ell,\slashed{p}_T) > 40~{\rm GeV}\, .
\label{cutsnlo}
\end{eqnarray}
We present these results in Table~\ref{tab:ktila_lhc_ol_lo} of the Appendix. 
Our results for $\tilde\kappa_\gamma$ can be approximated by
\begin{eqnarray}
\sigma(pp \to W^+\gamma) &=& \begin{cases} \left( 3860 +2434 \tilde\kappa_\gamma^2\right)~{\rm fb} &{\rm LO} \\
\left(5460+3241 \tilde\kappa_\gamma^2\right)~{\rm fb} & {\rm NLO} \end{cases} \nonumber \\
 \sigma(pp \to W^-\gamma)  &=& \begin{cases} \left(2878+1389 \tilde\kappa_\gamma^2\right)~{\rm fb} & {\rm LO} \\
\left(4100+1733 \tilde\kappa_\gamma^2\right)~{\rm fb} & {\rm NLO} \end{cases} .
\end{eqnarray}
The SM total cross sections are significantly increased at NLO, but interestingly the sensitivity to $\tilde\kappa_\gamma$ is not affected much, at 95\% c.l. we obtain from $\sigma(W^+\gamma)$ and $\sigma(W^-\gamma)$ respectively,
\begin{eqnarray}
\mid \tilde\kappa_\gamma \mid \lsim \begin{cases} 0.13 {\rm ~or~} 0.15 & {\rm LO}  \\
0.12 {\rm ~or~} 0.15 & {\rm NLO} \end{cases}
\end{eqnarray}

The asymmetry ${\cal A}_\ell$, however,  is only slightly changed at NLO,
\begin{eqnarray}
{\cal A}_{\ell}(pp \to W^+\gamma) &=& \begin{cases} -352\ \tilde\kappa_\gamma~{\rm fb} & {\rm LO} \\
-440\ \tilde\kappa_\gamma~{\rm fb} & {\rm NLO} \end{cases} \nonumber \\
 {\cal A}_{\ell}(pp \to W^-\gamma) &=& \begin{cases}-530\ \tilde\kappa_\gamma~{\rm fb} & {\rm LO} \\
-525\ \tilde\kappa_\gamma~{\rm fb} & {\rm NLO} \end{cases} 
\end{eqnarray}
The different cases, $W^\pm$ and LO vs NLO, give slightly different constraints but similar to those from the cross-sections.

\section{$pp \to Z\gamma \to \ell^\pm \ell^\mp \gamma$}

On-shell $Z\gamma$ production via a virtual boson $V=\gamma,Z$ can be described by the most general Lorentz and gauge 
invariant  anomalous $Z\gamma V$ 
coupling~\cite{Gounaris:2000tb},
\begin{eqnarray}
\Gamma^{\alpha\beta\mu}_{Z\gamma V}(q_1,q_2,P)&=&\frac{P^2-M_V^2}{ M_Z^2}
\left( h_1^V\left(q_2^\mu g^{\alpha\beta}-q_2^\alpha g^{\mu\beta}\right)+\frac{h_2^V}{M_Z^2}P^\alpha\left(P\cdot q_2g^{\mu\beta}-q_2^\mu P^\beta\right) \right.\nonumber \\
&+& \left. h_3^V \epsilon^{\mu\alpha\beta\rho}q_{2\rho}-\frac{h_4^V}{M_Z^2}P^\alpha\epsilon^{\mu\beta\rho\sigma}P_\rho q_{2\sigma}\right)\, ,
\end{eqnarray}
where $h^V_{1,2}$ are $CP$-odd and $h^V_{3,4}$ are $CP$-even. The factor $(P^2-M_V^2)$ is required by gauge invariance for 
$V=\gamma$ and by Bose symmetry for $V=Z$.  Limits on the anomalous $ZZ\gamma$ couplings have been obtained at the Tevatron~\cite{tevlimitZ},
as well as at the LHC~\cite{Aad:2012mr, Aad:2011tc,Martelli:2012np, Chatrchyan:2011rr}. For example, using cross-section measurements with 1.02 fb$^{-1}$ at 7~TeV, ATLAS finds the 95\% c.l. limits \cite{Aad:2011tc} 
\begin{eqnarray}
-0.028 \lsim h^\gamma_3 \lsim 0.027, && -0.022 \lsim h^Z_3 \lsim 0.026, \nonumber \\
 -0.00021 \lsim h^\gamma_4 \lsim 0.00021, && -0.00022 \lsim h^Z_4 \lsim 0.00021. 
\end{eqnarray}

Unlike the $W\gamma$ channel of the previous section, the $Z\gamma$ channel allows us to construct true CP odd observables. This happens because the final state in the decay mode $Z\gamma \to \ell^\pm \ell^\mp \gamma$ has definite CP properties that suffice to isolate CP violation. We will use the following observable
\begin{eqnarray}
{\cal O}_Z =  {\vec p}_{beam} . \left({\vec p}_{l^+}  \times {\vec p}_{l^-}\right)  {\vec p}_{beam} . \left({\vec p}_{l^+}  - {\vec p}_{l^-}\right), 
\label{tprodZ}
\end{eqnarray}
which generalizes the operator used in Ref.~\cite{Dawson:1996ge} and is quadratic in the beam momentum. This operator is also the same one used by Ref.~\cite{Han:2009ra} for the $pp\to W^+W^-$ process.  Based on Eq.~\ref{tprodZ} we construct the integrated asymmetry
\begin{equation}
{\cal A}_Z  \equiv \sigma({\cal O}_{Z} >0)-\sigma({\cal O}_{Z} <0)\, ,
\label{zasy}
\end{equation}
and define
\begin{equation}
\sigma_{Z\gamma}^+\equiv \sigma({\cal{O}}_Z> 0),\quad {\hbox{etc}}\, 
\end{equation}
for entries into the Table in the Appendix.

We  implement the CP violating couplings $h^1_\gamma$ and $h^1_Z$ into {\tt MadGraph 5} \cite{madgraph} with the aid of {\tt FeynRules} \cite{Christensen:2008py}.  
We use {\tt CTEQ6L1} pdfs evaluated at a scale $\mu_R = \mu_F = M_W$ with 
cuts $p_{T_l} > 20$ GeV, $p_{T_\gamma} > 10$ GeV, $\left|\eta_l\right| < 2.5$, $\left|\eta_\gamma\right| < 2.5$, 
$\Delta R_{l\gamma} > 0.7$, $\slashed p_T > 25$ GeV, and $M_T(l, \slashed p_T) > 40$ GeV to generate 
$10^6$ events  for the 14 TeV LHC. We reproduce our numerical results in Table~\ref{tab:h1_lhc_oz} for two different (large) values of the anomalous couplings as well as for the SM. 
Our numerical results can be summarized approximately by
\begin{eqnarray}
\sigma_{Z\gamma} &=& \left(475 + 0.032~\left(h^1_\gamma\right)^2 + 0.007~\left(h^1_Z\right)^2 \right){\rm fb} \nonumber \\
{\cal A}_Z &=& 0.1~h^1_\gamma - 0.013~h^1_Z \, .
\end{eqnarray}

From these expressions we estimate the $95\%$ c.l. limits for 10~fb$^{-1}$ at $\sqrt{S}$=14 TeV from measurements of the total
cross section and the asymmetry defined in Eq. \ref{zasy}  to be, 
\begin{eqnarray}
 {\rm {At}}~95\%~CL{\hbox{ and~ LO~ QCD}}:\quad
 |h^1_\gamma| &\lsim& \begin{cases} 21 & {\rm~from~} \sigma_{Z\gamma} \\
135 & {\rm ~from~} {\cal A}_Z \end{cases} \nonumber \\
|h^1_Z| &\lsim& \begin{cases} 44 & {\rm~from~} \sigma_{Z\gamma} \\
1039 & {\rm ~from~} {\cal A}_Z \end{cases}\, .
\label{zgam2}
\end{eqnarray}
The total cross-sections are much more sensitive to the presence of the CP violating anomalous couplings (through their quadratic effect) than is the CP-odd asymmetry. This implies that the bounds we are able to place do not single out the CP odd nature of the anomalous couplings.

To investigate the role of higher order QCD corrections, we use
{\tt MCFM 6.2} \cite{mcfm} to find $\sigma^{\rm NLO}_{SM} = 648.7$ fb which changes the limits by about 20\%. Given that these couplings, unlike $\tilde\kappa_\gamma$, only occur in the effective Lagrangian through operators  of dimension eight or more \cite{Dawson:1996ge} they do not merit further study.

\section{Conclusions}

We have estimated the sensitivity of the LHC to the  CP violating anomalous coupling $\tilde\kappa_\gamma$ and found it to be comparable to what ATLAS and CMS have extracted for its CP conserving counterpart $\kappa_\gamma$. We have presented results for the 7 and 14 TeV LHC using $W^\pm \gamma$ cross sections as well as T-odd asymmetries as observables. We find that the T-odd asymmetries offer a slightly higher sensitivity to $\tilde\kappa_\gamma$  than the cross sections. However, they do not single out this coupling as they are not true CP-odd observables. Comparing the results from {\tt MadGraph} and {\tt MCFM}, we find some uncertainty, of order 5\% for cross-sections and as large as 20\% for asymmetries, in the estimated sensitivities. We also estimated the effect on the sensitivity to the anomalous couplings due to NLO corrections  by generating events with {\tt MCFM} at both LO and NLO. Although there is some difference between the two cases, these differences amount to factors of two.

For the case of  $Z\gamma$ production we estimated the sensitivity to the CP violating couplings $h^1_{\gamma,Z}$ and found that in this case it is much worse than the corresponding sensitivity to CP conserving anomalous couplings. Although it is possible to construct true CP-odd observables in this case, they are not very sensitive to these anomalous couplings. Since these couplings originate at higher order in the effective Lagrangian for beyond SM physics (at least dimension eight operators) we do not pursue in detail effects of NLO corrections.

\begin{acknowledgments}

The work of G.V. (S.D.)  was supported in part by DOE under contract number
DE-FG02-01ER41155 (DE-AC02-98CH10886). The work of S. K. G. was supported in part by the ARC Centre of Excellence for Particle Physics at the Tera-scale. The use of Monash University Sun Grid, a high-performance computing 
facility, is gratefully acknowledged.  We thank A. Goshaw for useful discussions.

\end{acknowledgments}
\newpage
\appendix

\section{Tables for the $W\gamma$ process}

\begin{table}[h]
\caption{\sf\small Tevatron analysis of the asymmetry defined in Eq. \ref{astriprods} using 
 ${\cal O}_W$ with non-zero $\kappa_\gamma$ and $\tilde\kappa_\gamma$ 
as defined in Eq.  \ref{hagiwwvdef} for the process $p{\overline p}\rightarrow
W^\pm \gamma \to l^\pm \nu_l \gamma$. The input parameters and cuts are discussed in the text. All the numbers shown here are in fb units.}
\vskip .25in
\centering
\resizebox{12.5cm}{!} {
\begin{tabular}{|c|r@{$\pm$}l|r@{$\pm$}l|r@{$\pm$}l|r@{$\pm$}l|r@{$\pm$}l|r@{$\pm$}l|}\hline\hline
& \multicolumn{4}{|c|}{SM} & \multicolumn{4}{|c|}{NP1 (${\tilde\kappa}_\gamma = 0.3, \kappa_\gamma = 0$)}
& \multicolumn{4}{|c|}{NP2 (${\tilde\kappa}_\gamma = 0, \kappa_\gamma = 0.3$)}\\\hline
& \multicolumn{2}{|c|}{$W^+$}& \multicolumn{2}{|c|}{$W^-$} & \multicolumn{2}{|c|}{$W^+$} & \multicolumn{2}{|c|}{$W^-$} 
& \multicolumn{2}{|c|}{$W^+$} & \multicolumn{2}{|c|}{$W^-$} \\\hline\hline
$\sigma^+ $ &  99.56 & 0.06 &  99.82 & 0.06  &  98.25 & 0.06 & 145.98 & 0.06  & 127.32 & 0.22 & 127.12 & 0.22 \\\hline
$\sigma^- $ & 100.57 & 0.06 & 100.20 & 0.06  & 145.46 & 0.06 &  98.31 & 0.06  & 126.49 & 0.21 & 126.55 & 0.21 \\\hline
$\sigma   $ & 200.13 & 0.08 & 200.02 & 0.08  & 243.71 & 0.08 & 244.29 & 0.08  & 253.81 & 0.30 & 253.67 & 0.30 \\\hline
${\cal A}_W $ &  -1.01 & 0.08 &  -0.38 & 0.08  & -47.21 & 0.08 &  47.67 & 0.08  &   0.83 & 0.30 &   0.57 & 0.30 \\\hline
\end{tabular}
}
\label{tab:ktila_tev_ow}
\end{table}

\begin{table}[h]
\caption{\sf\small Results for the asymmetries defined in Eq. \ref{astriprods} 
in the process  $p p\to W^\pm \gamma \to \ell^\pm \nu \gamma$ at $\sqrt{S}= 14$ TeV for non-zero
${\tilde\kappa}_\gamma$ and
${\kappa}_\gamma$. The input parameters and
cuts are discussed in the text and these results are obtained with {\tt MadGraph}. All the numbers shown here are in fb units.}
\vskip .25in
\centering
\resizebox{16.0cm}{!} 
{
\begin{tabular}{|c|r@{$\pm$}l|r@{$\pm$}l|r@{$\pm$}l|r@{$\pm$}l|r@{$\pm$}l|r@{$\pm$}l|r@{$\pm$}l|r@{$\pm$}l|r@{$\pm$}l|r@{$\pm$}l|}\hline\hline
& \multicolumn{4}{|c|}{SM} 
& \multicolumn{4}{|c|}{NP1 (${\tilde\kappa}_\gamma = 0.3, \kappa_\gamma = 0$)}
& \multicolumn{4}{|c|}{NP1 (${\tilde\kappa}_\gamma = 0.1, \kappa_\gamma = 0$)}
& \multicolumn{4}{|c|}{NP2 (${\tilde\kappa}_\gamma = 0, \kappa_\gamma = 0.3$)}
& \multicolumn{4}{|c|}{NP2 (${\tilde\kappa}_\gamma = 0, \kappa_\gamma = 0.1$)}
\\\hline
& \multicolumn{2}{|c|}{$W^+$} & \multicolumn{2}{|c|}{$W^-$}
& \multicolumn{2}{|c|}{$W^+$} & \multicolumn{2}{|c|}{$W^-$}
& \multicolumn{2}{|c|}{$W^+$} & \multicolumn{2}{|c|}{$W^-$}
& \multicolumn{2}{|c|}{$W^+$} & \multicolumn{2}{|c|}{$W^-$} 
& \multicolumn{2}{|c|}{$W^+$} & \multicolumn{2}{|c|}{$W^-$}
\\\hline\hline
$\sigma          $ & 659.6 & 0.9 & 549.6 & 0.8 & 1070.8 & 1.1 & 823.8  & 1.1  &705.2 & 0.8 &579.7& 0.9 &1105.9 & 1.2 & 838.3 & 1.2 & 717.2&1.0&586.1&0.8\\\hline
${\cal A}_W      $ &  -0.4 & 0.8 &   0.0 & 0.7 &    1.1 & 1.0 &    0.0 & 1.2 &   0.1 & 0.8 &   0.0 & 0.8 &  0.0 & 1.3 &  0.0 & 1.1 &  0.0 & 0.9 &  0.0 & 0.8 \\\hline
${\cal A}_\gamma $ &  -3.5 & 0.9 &   4.2 & 0.8 &  138.2 & 1.0 &  136.0 & 1.1 &  43.9 & 0.8 &  47.7 & 0.9 & -0.6 & 1.2 &  3.6 & 1.2 & -2.0 & 1.0 &  2.8 & 0.8\\\hline
${\cal A}_\ell   $ &   5.8 & 0.8 &  -3.5 & 0.8 & -129.5 & 1.1 & -126.4 & 1.1 & -41.6 & 0.8 & -44.3 & 0.8 &  3.0 & 1.2 & -2.7 & 1.1 &  3.5 & 0.9 & -2.5 & 0.8\\\hline
\end{tabular}
}
\label{tab:ktila_lhc_ow}
\end{table}

\begin{table}[h]
\caption{\sf\small Results (in fb units) for the asymmetries defined in Eq. \ref{astriprods} for the process $p{p}\to W^\pm \gamma \to l^\pm \nu_l \gamma$ at $\sqrt{S}=
14$ TeV using {\tt MCFM}. Parton densities ({\tt CTEQ6L1} for LO and {\tt CTEQ6.6M} for NLO) are evaluated at a scale $\mu_R = \mu_F = M_W$. The input parameters and cuts are discussed in the text. All the numbers shown here are in fb units.}\vskip .25in
\centering
\resizebox{16.0cm}{!}
{
\begin{tabular}{|c|r@{$\pm$}l|r@{$\pm$}l|r@{$\pm$}l|r@{$\pm$}l|r@{$\pm$}l|r@{$\pm$}l|r@{$\pm$}l|r@{$\pm$}l|r@{$\pm$}l|r@{$\pm$}l|}\hline\hline
& \multicolumn{4}{|c|}{SM}
& \multicolumn{4}{|c|}{NP1 (${\tilde\kappa}_\gamma = 0.3, \kappa_\gamma = 0$)}
& \multicolumn{4}{|c|}{NP1 (${\tilde\kappa}_\gamma = 0.1, \kappa_\gamma = 0$)}
& \multicolumn{4}{|c|}{NP2 (${\tilde\kappa}_\gamma = 0, \kappa_\gamma = 0.3$)}
& \multicolumn{4}{|c|}{NP2 (${\tilde\kappa}_\gamma = 0, \kappa_\gamma = 0.1$)}
\\\hline
& \multicolumn{2}{|c|}{$W^+$} & \multicolumn{2}{|c|}{$W^-$}
& \multicolumn{2}{|c|}{$W^+$} & \multicolumn{2}{|c|}{$W^-$}
& \multicolumn{2}{|c|}{$W^+$} & \multicolumn{2}{|c|}{$W^-$}
& \multicolumn{2}{|c|}{$W^+$} & \multicolumn{2}{|c|}{$W^-$}
& \multicolumn{2}{|c|}{$W^+$} & \multicolumn{2}{|c|}{$W^-$}
\\\hline\hline
& \multicolumn{20}{|c|}{LO} \\\hline\hline
$\sigma          $ &686.3&0.4 &570.4&0.3 &1044.7&0.5 &818.0&0.4 &726.0&0.4 &598.1&0.3 &1009.3&0.5 &826.0&0.4 &714.4&0.4 &600.9&0.3\\\hline
${\cal A}_\ell   $ &0.5&0.4 &-1.9&0.3 &-130.7&0.5 &-166.9&0.4 &-43.1&0.4 &-56.9&0.3 &1.8&0.5 &-2.4&0.4 &0.6&0.4 &-1.9&0.3\\\hline\hline
& \multicolumn{20}{|c|}{NLO} \\\hline\hline
$\sigma          $ &1704.8&0.8 &1456.5&0.7 &2200.6&1.0 &1774.9&0.8 &1763.9&0.9 &1489.1&0.7 &2269.1&1.1 &1702.4&0.8 &1786.8&0.9 &1464.8&0.7\\\hline
${\cal A}_\ell   $ &-9.4&0.8 &9.1&0.7 &-166.8&1.0 &-181.7&0.8 &-61.9&0.9 &-54.6&0.7 &-26.2&1.1 &-8.8&0.8 &-14.8&0.9 &2.8&0.7\\\hline\hline
\end{tabular}
}
\label{tab:mcfm_wga_lhc_14}
\end{table}

The results of the {\tt MCFM} runs at 14 TeV can be approximated with the following fits, to be compared to Eq.~\ref{wgres} for {\tt MadGraph}.
\begin{eqnarray}
\sigma(W^+\gamma) &\approx & \left( 686 - 117 ~\kappa_\gamma + 3978~ \kappa^2_\gamma -  1.8 ~{\tilde\kappa}_\gamma +3988 ~ {\tilde \kappa}^2_\gamma\right) {\rm ~fb}\nonumber \\
\sigma(W^-\gamma) &\approx & \left(570 + ~32 ~\kappa_\gamma + 2735~ \kappa^2_\gamma +2.8 ~{\tilde\kappa}_\gamma + 2742~ {\tilde \kappa}^2_\gamma\right) {\rm ~fb}\nonumber \\
{\cal A}_\gamma (W^+\gamma) &=& -{\cal A}_\ell(W^+\gamma)  \ \approx \ 436 \tilde\kappa_\gamma
\nonumber \\
{\cal A}_\gamma (W^-\gamma) &=& -{\cal A}_\ell(W^-\gamma)  \ \approx \ 555 \tilde\kappa_\gamma \, .
\label{wgresmcfmlo}
\end{eqnarray}
and at NLO  
\begin{eqnarray}
\sigma(W^+\gamma) &\approx & \left( 1705 +290 ~\kappa_\gamma + 5303~ \kappa^2_\gamma +60 ~{\tilde\kappa}_\gamma +5308 ~ {\tilde \kappa}^2_\gamma\right) {\rm ~fb}\nonumber \\
\sigma(W^-\gamma) &\approx & \left(1457 -285 ~\kappa_\gamma + 3683~ \kappa^2_\gamma -42 ~{\tilde\kappa}_\gamma + 3677~ {\tilde \kappa}^2_\gamma\right) {\rm ~fb}\nonumber \\
{\cal A}_\gamma (W^+\gamma) &=& -{\cal A}_\ell(W^+\gamma)  \ \approx \ 552 \tilde\kappa_\gamma
\nonumber \\
{\cal A}_\gamma (W^-\gamma) &=& -{\cal A}_\ell(W^-\gamma)  \ \approx \ 610 \tilde\kappa_\gamma \, .
\label{wgresmcfmnlo}
\end{eqnarray}

\begin{table}[h] 
\caption{\sf\small Results (in fb units) for the asymmetries defined in Eq. \ref{astriprods} for the process $p{p}\to W^\pm \gamma \to l^\pm \nu_l \gamma$ at $\sqrt{S}= 
7$ TeV for non-zero ${\tilde\kappa}_\gamma$. Parton densities ({\tt CTEQ6L1} for LO and {\tt CTEQ6.6M} for NLO) are evaluated at a scale $\mu_R = \mu_F = M_W$. The cuts are: $p_{T_l} > 20$ GeV, 
$p_{T_\gamma} > 10$ GeV, $\left|\eta_l\right| < 2.5$, $\left|\eta_\gamma\right| < 2.5$, $\Delta R_{l\gamma} > 0.7$, $\slashed p_T > 25$ GeV, and $M_T(l, \slashed p_T) > 40$ GeV. } \vskip .25in

\centering
\resizebox{8.5cm}{!} {
\begin{tabular}{|c|r@{ $\pm$ }l|r@{ $\pm$ }l|r@{ $\pm$ }l|r@{ $\pm$ }l|}\hline\hline
& \multicolumn{4}{|c|}{SM} & \multicolumn{4}{|c|}{NP (${\tilde\kappa}_\gamma = 0.3$)}\\\hline\hline
& \multicolumn{2}{|c|}{$W^+$}& \multicolumn{2}{|c|}{$W^-$} & \multicolumn{2}{|c|}{$W^+$} & \multicolumn{2}{|c|}{$W^-$}\\\hline\hline
& \multicolumn{8}{|c|}{LO} \\\hline\hline
$\sigma^+ $ &  1928.1 & 3.8 &  1432.7 & 2.7 &  1985.7 & 3.8 &  1421.8 & 2.6\\\hline
$\sigma^- $ &  1929.7 & 3.7 &  1445.1 & 2.7 &  2091.2 & 4.1 &  1581.0 & 2.8\\\hline
$\sigma   $ &  3857.8 & 5.3 &  2877.8 & 3.8 &  4076.9 & 5.4 &  3002.9 & 3.8\\\hline
${\cal A}_{\ell} $ &    -1.6 & 5.3 &   -12.4 & 3.8 &  -105.5 & 5.4 &  -159.2 & 3.8\\\hline
\hline
& \multicolumn{8}{|c|}{NLO} \\\hline\hline
$\sigma^+ $ &  2728.4 & 6.5 & 2052.4 & 5.1 &  2809.4 & 6.9 &  2049.4 & 4.4\\\hline
$\sigma^- $ &  2730.7 & 6.5 & 2047.4 & 4.7 &  2941.5 & 6.6 &  2207.0 & 4.8\\\hline
$\sigma   $ &  5459.2 & 9.2 & 4099.8 & 6.9 &  5750.9 & 9.8 &  4256.3 & 6.5\\\hline
${\cal A}_{\ell} $ &    -2.3 & 9.2 &    5.0 & 6.9 &  -132.1 & 9.8 &  -157.6 & 6.5\\\hline
\end{tabular}
}
\label{tab:ktila_lhc_ol_lo}
\end{table}

\section{Tables for the $Z\gamma$ process}

\begin{table}[h]
\caption{\sf\small Results (in fb units) for the process $pp \to Z \gamma \to l^+ l^- \gamma$ at $\sqrt{S}= 14$ TeV for $h^\gamma_1$ and
$h^Z_1$ respectively. Parton densities ({\tt CTEQ6L1}) are evaluated at a scale $\mu_R = \mu_F = \sqrt{M^2_W + 
p^2_{T_\gamma}}$. The cuts are: $p_{T_{l^\pm}} > 20$ GeV, $p_{T_\gamma} > 20$ GeV, $\left|\eta_{l^\pm}\right| < 2.5$, 
$\left|\eta_\gamma\right| < 2.5$, $\Delta R_{l^+l^-} > 0.4$, $\Delta R_{l^\pm\gamma} > 0.7$, and, $M_{l^+l^- \gamma}> 100$ 
GeV. The operator used in this case is ${\cal O}_Z$.}
\vskip .25in
\centering
\resizebox{10.5cm}{!} {
\begin{tabular}{|c|r@{$\pm$}l|r@{$\pm$}l|r@{$\pm$}l|r@{$\pm$}l|r@{$\pm$}l|}\hline\hline
& \multicolumn{2}{|c|}{SM} & \multicolumn{4}{|c|}{NP1 ($h^\gamma_1$)}
& \multicolumn{4}{|c|}{NP2 ($h^Z_1$)}\\\hline
& \multicolumn{2}{|c|}{$0$} & \multicolumn{2}{|c|}{$100$} & \multicolumn{2}{|c|}{$500$}
& \multicolumn{2}{|c|}{$100$} & \multicolumn{2}{|c|}{$500$} \\\hline\hline
$\sigma^+_{Z\gamma} $  & 237.49 & 0.20  & 396.30 & 0.25 & 4285.81 & 3.07  & 269.94 & 0.26 & 1097.54 & 0.87 \\\hline
$\sigma^-_{Z\gamma} $  & 237.71 & 0.20  & 387.65 & 0.25 & 4234.50 & 3.04  & 272.0  & 0.26 & 1104.16 & 0.88 \\\hline
$\sigma_{Z\gamma}   $  & 475.20 & 0.28  & 783.95 & 0.35 & 8520.30 & 4.32  & 541.94 & 0.37 & 2201.7  & 1.23 \\\hline
${\cal A} $  &  -0.22 & 0.28  &   8.65 & 0.35 &   51.31 & 4.32  &  -2.06 & 0.37 &   -6.62 & 1.24 \\\hline
\end{tabular}
}
\label{tab:h1_lhc_oz}
\end{table}

\newpage



\newpage



\begin{thebibliography}{999}

\bibitem{ATLAS:2012mec} 
  G.~Aad {\it et al.}  [ATLAS Collaboration],
  arXiv:1210.2979 [hep-ex].


\bibitem{Aad:2012mr} 
  G.~Aad {\it et al.}  [ATLAS Collaboration],
  Phys.\ Lett.\ B {\bf 717}, 49 (2012)
  [arXiv:1205.2531 [hep-ex]].

\bibitem{Aad:2011tc} 
  G.~Aad {\it et al.}  [ATLAS Collaboration],
  JHEP {\bf 1109}, 072 (2011)
  [arXiv:1106.1592 [hep-ex]].
  
\bibitem{Martelli:2012np} 
  A.~Martelli {\it et al.}  [CMS Collaboration],
  EPJ Web Conf.\  {\bf 28}, 06002 (2012)
  [arXiv:1201.4596 [hep-ex]].

\bibitem{Chatrchyan:2011rr} 
  S.~Chatrchyan {\it et al.}  [CMS Collaboration],
  Phys.\ Lett.\ B {\bf 701}, 535 (2011)
  [arXiv:1105.2758 [hep-ex]].
  


\bibitem{Hagiwara:1986vm} 
  K.~Hagiwara, R.~D.~Peccei, D.~Zeppenfeld and K.~Hikasa,
  Nucl.\ Phys.\ B {\bf 282}, 253 (1987).


\bibitem{Gounaris:2000tb} 
  G.~J.~Gounaris, J.~Layssac and F.~M.~Renard,
  Phys.\ Rev.\ D {\bf 62}, 073013 (2000)
  [hep-ph/0003143].

\bibitem{Dawson:1996ge}
  S.~Dawson, X.~G.~He and G.~Valencia,
  Phys.\ Lett.\  B {\bf 390}, 431 (1997)
  [arXiv:hep-ph/9609523];

\bibitem{tprods}
  J.~F.~Donoghue and G.~Valencia,
  Phys.\ Rev.\ Lett.\  {\bf 58}, 451 (1987)
  [Erratum-ibid.\  {\bf 60}, 243 (1988)];
  
  M.~B.~Gavela, F.~Iddir, A.~Le Yaouanc, L.~Oliver, O.~Pene and J.~C.~Raynal,
  Phys.\ Rev.\  D {\bf 39}, 1870 (1989);
  M.~P.~Kamionkowski,
  Phys.\ Rev.\  D {\bf 41}, 1672 (1990).
  

\bibitem{tevlimitW}
  V.~M.~Abazov {\it et al.}  [D0 Collaboration],
  Phys.\ Rev.\ Lett.\  {\bf 107}, 241803 (2011)
  [arXiv:1109.4432 [hep-ex]].
  
\bibitem{Alcaraz:2006mx} 
  J.~Alcaraz {\it et al.}  [ALEPH and DELPHI and L3 and OPAL and LEP Electroweak Working Group Collaborations],
  hep-ex/0612034.

\bibitem{Baur:1989gk} 
  U.~Baur and E.~L.~Berger,
  Phys.\ Rev.\ D {\bf 41}, 1476 (1990).
  
\bibitem{Appelquist:1994qz} 
  T.~Appelquist and G.~-H.~Wu,
  Phys.\ Rev.\ D {\bf 51}, 240 (1995)
  [hep-ph/9406416].
  

\bibitem{Han:2009ra} 
  T.~Han and Y.~Li,
  Phys.\ Lett.\ B {\bf 683}, 278 (2010)
  [arXiv:0911.2933 [hep-ph]].

\bibitem{Kumar:2008ng} 
  J.~Kumar, A.~Rajaraman and J.~D.~Wells,
  Phys.\ Rev.\ D {\bf 78}, 035014 (2008)
  [arXiv:0801.2891 [hep-ph]].
  
  
  
  
  \bibitem{madgraph}
  T.~Stelzer and W.~F.~Long,
  Comput.\ Phys.\ Commun.\  {\bf 81}, 357 (1994)
  [arXiv:hep-ph/9401258];
  J.~Alwall {\it et al.},
  JHEP {\bf 0709}, 028 (2007)
  [arXiv:0706.2334 [hep-ph]];
  J.~Alwall, M.~Herquet, F.~Maltoni, O.~Mattelaer, T.~Stelzer,
  JHEP {\bf 1106}, 128 (2011).
  [arXiv:1106.0522 [hep-ph]].

\bibitem{Christensen:2008py} 
  N.~D.~Christensen and C.~Duhr,
  Comput.\ Phys.\ Commun.\  {\bf 180}, 1614 (2009)
  [arXiv:0806.4194 [hep-ph]].
  
  
\bibitem{Marciano:1986eh} 
  W.~J.~Marciano and A.~Queijeiro,
  Phys.\ Rev.\ D {\bf 33}, 3449 (1986).
  

  
\bibitem{Baur:1993ir} 
  U.~Baur, T.~Han and J.~Ohnemus,
  Phys.\ Rev.\ D {\bf 48}, 5140 (1993)
  [hep-ph/9305314].

\bibitem{mcfm} 
  J.~M.~Campbell and R.~K.~Ellis,
  Nucl.\ Phys.\ Proc.\ Suppl.\  {\bf 205-206}, 10 (2010)
  [arXiv:1007.3492 [hep-ph]].

\bibitem{cmswgstudy}
See for example, {\tt https://twiki.cern.ch/twiki/bin/view/Main/SyueWeiLiWgammaNLOCrossSection}
  

 \bibitem{tevlimitZ} 
  V.~M.~Abazov {\it et al.}  [D0 Collaboration],
  Phys.\ Rev.\ D {\bf 85}, 052001 (2012)
  [arXiv:1111.3684 [hep-ex]];
  T.~Aaltonen {\it et al.}  [CDF Collaboration],
  Phys.\ Rev.\ Lett.\  {\bf 107}, 051802 (2011)
  [arXiv:1103.2990 [hep-ex]].
  

\end{thebibliography}
\end{document}